\documentclass{article}
\usepackage{fleqn,epsfig}
\usepackage{amsmath,amssymb}
\newcommand{\bc}{\begin{center}}
\newcommand{\ec}{\end{center}}
\newcommand{\be}{\begin{equation}}
\newcommand{\ee}{\end{equation}}
\newcommand{\bea}{\begin{eqnarray}}
\newcommand{\eea}{\end{eqnarray}}
\newcommand{\bsl}{\boldsymbol}
\begin{document}
\begin{center}
{\Large\bf \boldmath Baryons in the
Field Correlator Method:\\[1mm] Effects of the Running Strong Coupling } 

\vspace*{6mm}
{R. Ya. Kezerashvili$^{1}$, I. M. Narodetskii$^{1,2}$, A. I. Veselov$^{2}$ }\\[5mm]
$^{1}$Physics Department, The City University of New York, New York, 11201, USA\\[2mm]
$^{2}$Institute of Theoretical and Experimental Physics, Moscow
117218, Russia\\[2mm]
\end{center}

\vspace*{6mm}

\begin{abstract}
The ground and $P$-wave excited states of $nnn$, $nns$ and $ssn$
baryons are studied in the framework of the field correlator
method using the  running strong coupling constant in the
Coulomb--like part of the three--quark potential. The running
coupling is calculated up to two loops in the background
perturbation theory. The three-quark problem has been solved using
the hyperspherical functions method. The masses of the $S$-- and
$P$--wave baryons  are presented. Our approach reproduces and
improves the previous results for the baryon masses obtained for
the freezing value of the coupling constant. The string correction
for the confinement potential of the orbitally excited baryons,
which is the leading contribution of the proper inertia of the
rotating strings, is estimated.

\end{abstract}
\section{Introduction}
Quantum chromodynamics (QCD) has been established as the theory
describing the strong interaction but its application to
low--energy hadron phenomenology is still far from a routine
deduction. Various approximations, whose connection to the
underlying theory remains sometimes obscure, are presently used to
describe baryon spectroscopy. There is a vast literature of
theoretical treatment of baryons, including treatment in specific
quark models \cite{PDG:review}, QCD sum rules \cite{QCDSR},
effective field theories \cite{EFT},  the Skyrme
model \cite{Oh07}, the collective models of the baryon such as the
algebraic approach of Ref. \cite{BIL00}, and large $N_c$ analysis
\cite{CC00}, as well as on the lattice \cite{Lattice}. It has
become an attractive program to develop model independent methods
which are firmly based in fundamental theory. In particular,
ground state spectroscopy in lattice QCD calculations appears to
be well understood \cite{Richards2008}. Excited state spectroscopy
on the lattice, however, is still a challenging task \cite{B06}.

The field correlator method (FCM) in QCD \cite{DS} provides
another perspective. The FCM is a promising formulation of the
nonperturbative QCD that gives additional support to the quark
model assumptions. Progress was recently made
\cite{NT2004}--\cite{NSV} towards placing the computation of
baryon masses within the FCM on the same level as that of mesons
\cite{BB2002} . Nevertheless, this work can be refined. In Refs.
\cite{NT2004}--\cite{NSV} a freezing value of the strong coupling
constant in the perturbative Coulomb--like potential has been
employed. This choice appears to be a reasonable approximation and
gives rise to a good description of heavy quarkonia \cite{E75} and
heavy--light mesons \cite{KNS}. Note that within QCD most of the
baryon mass is produced by quark confinement. For light baryons
the Coulomb--like force does not play a crucial role and produces
only a marginal ($\sim 10\%$) correction \cite{NT2004}.
Nevertheless, it is important to include into the FCM
approach the modern knowledge about the one--gluon exchange
forces that still represent a fundamental concept which might give
us a deeper understanding of baryon spectroscopy \cite{JPS1998}.

The objective of this paper is to explore the FCM for baryons in a
more regular way by considering the effects of the running strong
coupling constant in the Coulomb--like part of the three--quark
potential. We use the background perturbation theory (BPTh)
\cite{Simonov1995} for the coupling constant to avoid the infrared
singularities.  Below are considered the baryons composed of
valence light quarks, the {\it up, down,} and {\it strange}
flavors. We present the new results for the masses of the ground
states and $P$--wave excited states of $nnn$, $nns$ and $ssn$
baryons. We also for the first time estimate the string correction
to the baryonic orbital excitations. So far, this correction was
calculated for the orbitally excited mesons \cite{DKS}, \cite{KNS}
and hybrid charmonium states \cite{KN2008}.

This paper is organized as follows. In Sec. 2, we give a brief
summary of the effective Hamiltonian (EH) method important for our
particular calculation\footnote{The application of this method for
the baryons was described in detail elsewhere \cite{NT2004, DNV,
NSV}.}.
 In Sec. 3, we discuss the calculation of the string correction. In Sect. 4,
 we use BPTh for the strong
coupling constant to define the Coulomb--like potentials. A brief
description of
the hyperspherical approach, which is a very effective numerical
tool for solving the three--quark problem is given in Sect. 5. Our
results for the masses of the $nnn$, $nns$ and $ssn$ baryons
including comparison to the previous ones obtained within the FCM
with a freezing value of the strong coupling constant are
presented in Sec. 6. The concluding remarks  follow  in Sec. 7.

\section{Effective Hamiltonian in FCM} In the FCM three--quark dynamics
in a baryon is encoded in gluonic field correlators which are
responsible for quark confinement. Starting from the
Feynman--Schwinger representation for the quark and gluon
propagators in the external field, one can extract hadronic
Green's functions and calculate the baryon spectra. The QCD string
model corresponds to the limit of a small gluonic correlation
length of field correlators.

The key ingredient of the FCM is the use of the auxiliary fields
(AF) initially introduced in order to get rid of the square roots
appearing in the relativistic Hamiltonian. Historically the AF
formalism was first introduced in Ref. \cite{P} to treat the kinematics
of the relativistic spinless particles. Using the AF formalism
allows one to derive a simple local form of the EH for the three-quark system \footnote{See Ref. \cite{Si2003}; for a brief review
of the EH formalism relevant to the baryon problem considered in
this paper see Sec. II of Ref. \cite{NT2004}}, which comprises
both confinement and relativistic effects, and contains only
universal parameters: the string tension $\sigma$, the strong
coupling constant $\alpha_s$, and the bare (current) quark masses
$m_i$. In this paper, we do not consider the spin--dependent part
of the EH\footnote{The spin-dependent effects for

baryons within the FCM will be considered elsewhere. For the

recent review of spin--dependent interactions in quarkonia in the framework the FCM see \cite{BNS08}.}.
Then the EH has the form
\begin{equation}
\label{eq:H} H=\sum\limits_{i=1}^3\left(\frac {m_{i}^2}{2\,\mu_i}+
\frac{\mu_i}{2}\right)+H_0+V.
\end{equation}
In Eq. (\ref{eq:H}) $H_0$ is the nonrelativistic kinetic energy
operator for the constant  AF $\mu_i$, the spin-independent
potential $V$ is the sum of the string potential
\be\label{eq:string} V_Y({\bf r}_1,\,{\bf r}_2,\,{\bf
r}_3)\,=\,\sigma\,r_{min},\ee with $r_{min}$ being the minimal
string length corresponding to the Y--shaped configuration, and a
Coulomb interaction term \be\label{V_C} V_{\rm C}({\bf r}_1,\,{\bf
r}_2,\,{\bf r}_3)\,=\,\sum_{i<j}\,V_C(r_{ij}),\ee  arising from
the one-gluon exchange.

The EH depends explicitly on both bare quark masses $m_i$ and the
constants AF $\mu_i$ that finally acquire the meaning of the
dynamical quark masses. As the first step the eigenvalue problem
is solved for each set of $\mu_i$; then one has to minimize
$\langle H\rangle$ with respect to $\mu_i$.
Although being formally simpler the EH  is equivalent to the
relativistic Hamiltonian up to elimination of AF \cite{NSV}.

 The formalism allows for a very
transparent interpretation of AF $\mu_i$: starting from bare quark
masses $m_i$, we arrive at the dynamical masses $\mu_i$ that
appear due to the interaction and can be treated as the dynamical
masses of constituent quarks. Therefore the constituent quark
masses appear in the FCM calculations. These have obvious quark
model analogs, but are derived directly using the AF formalism.
Due to confinement $\mu_i\,\sim\, \sqrt{\sigma}\,\sim\,400$ MeV or
higher, even for the massless current quarks.

The baryon mass is given by
\begin{equation}
\label{M_B} M_B\,=\,M_0\,+\,\Delta M_{\rm
string}\,+\,C,\end{equation}
\begin{equation}
\label{eq:M_B0}M_0\,=\,\sum\limits_{i=1}^3\left(\frac
{m_{i}^2}{2\mu_i\,}+
\,\frac{\mu_i}{2}\right)\,+\,E_0(\mu_i)\end{equation} where
$E_0(\mu_i)$ is an eigenvalue of the Schr\"{o}dinger operator $H_0
+V$, and the $\mu_i$ are defined by minimization condition
\be\label{eq:mc}
\frac{\partial\,M_0(m_i,\mu_i)}{\partial\,\mu_i}\, =\,0.
\end{equation} The right--hand side of Eq. (\ref{M_B}) contains
the perturbative quark self-energy correction $C$ that is created
by the color magnetic moment of a quark propagating through the
vacuum background field \cite{S2001}. This correction
adds an overall negative constant to the
hadron masses:
\begin{equation} \label{self_energy}
C\,=\,-\frac{2\sigma}{\pi}\,\sum\limits_i\frac{\eta(t_i)}{\mu_i},
\,\,\,\,\,t_i\,=\,m_i/T_g,\end{equation} where $1/T_g$ is the
above mentioned gluonic correlation length. The numerical factor
$\eta(t)$ arises from the evaluation of the integral
\be \eta(t)= t\int^\infty_0 z^2\, K_1(tz)\,
e^{-z}\,dz,\label{eq:eta} \ee where $K_1$ is the McDonald
function.
In what follows we use $T_g\,=\, 1$ GeV. Finally, $\Delta M_{\rm
string}$ in Eq. (\ref{M_B}) is the so--called string correction to
be discussed in the next section.

The above ingredients provide the basic mechanism that determines
the baryon spectrum in the FCM.
\section{The string
correction} The string potential $V_Y({\bf r}_1,\,{\bf r}_2,\,{\bf
r}_3)$ in Eq. (\ref{eq:string}) represents only the leading term
in the expansion of the QCD string Hamiltonian in terms of angular
velocities \cite{DKS}. The leading correction in this expansion is
known as a string correction. This is the correction totally
missing in relativistic equations with local potentials. Its sign
is negative, so the contribution of the string correction lowers
the energy of the system, thus giving a negative contribution to
the masses of orbitally excited states, leaving the S--wave states
intact.

As was mentioned in the introduction, so far, the string
correction was calculated only for the orbitally excited
heavy--light mesons \cite{KNS} and hybrid charmonium states
\cite{KN2008}. For a baryon with the genuine string junction point
the calculation of the string correction is a very cumbersome
problem. The calculations are greatly simplified, however,  if the
string junction point is chosen to coincide with the
center--of--mass coordinate ${\bsl R}_{\text cm}$. In this case,
the complicated string junction potential is approximated by a sum
of the one--body confining potentials. The accuracy of this
 approximation for the $P$--wave baryon states
 is better than 1$\%$ \cite{NSV}.
 Letting ${\bsl R}_{cm}\,=\,0$ we arrive at the following form
of the string potential
 \be\label{eq:V_string} V_{\rm string}^{\rm CM}\,=\,\sigma \,\sum_i\,|\bsl{r}_i|\int^1_0 d\beta\sqrt{1-{\bsl l}_i^2}
, \ee where \be{\bsl l}_i\,=\,\frac{\beta}{|{\bsl r}_i|}\,[{\bsl
r}_i\,\times\,\dot{\bsl r}_i]\,=\,\frac{\beta}{\mu_i\,|{\bsl
r}_i|}\,[{\bsl r}_i\,\times\,{\bsl
p}_i]\,=\,-i\,\frac{\beta}{\mu_i\,|{\bsl r}_i|}\, [{\bsl
r}_i\,\times\,{\bsl\nabla}_i].\ee Expanding the square roots in
Eq. (\ref{eq:V_string}) in powers of the angular velocities ${\bsl
l}_i^2$ and keeping only the first two terms in this expansion,
one obtains
 \be V_{\rm
string}^{\rm CM}\,\approx\,\,\sigma\,\sum_i\,|{\bsl
r}_i|\,\int\limits_0^1\,d\beta\,(1\,-\,\frac{1}{2}\,{\bsl
l}_i^2)\,=\,\sigma\,\sum_i\,|{\bsl
r}_i|\,+\,\frac{\sigma}{6}\,\sum_i\left(\frac{1}{\mu_i^2\,|{\bsl
r}_i|}({\bsl r}_i\,\times\,{\bsl\nabla}_i)^2\right) \ee and
 \be\label{eq:Delta_M}\Delta M_{\rm
string}\,=\,-\,\frac{\sigma}{6}\,<\Psi|\sum_i\,\frac{({\bsl
r}_i\times{\bsl p}_i)^2}{\mu_i^2\,r_i}\,|\Psi>,\ee where $\Psi$ is
an eigenfunction of the Hamiltonian (\ref{eq:H}).  Further details
and numerical results are  given in Sec. \ref{sect:results}.

\section{Coulomb--like interaction in the BPTh}
As was mentioned above, the spin--independent potential in Eq.
(\ref{eq:H}) is the sum of the confinement potential and
Coulomb--like interaction. Details of our treatment of the string
junction confinement potential can be found in the appendix of
Ref. \cite{DNV}. Let us now concentrate on the Coulomb--like part
of interaction, $V_C(r)$. It is convenient to write the
Coulomb--like  potential in QCD in momentum space as
\be\label{eq:MS} V_C({\textbf
q}^2)\,=\,-\,C_F\,\frac{\alpha_V({\textbf q}^2)}{{\textbf
q}^2},\ee where $C_F$ is the color factor. In a baryon, the two
quarks belong to the representation ${\bf\overline 3}$ of
$SU_c(3)$ for which $C_F\,=\,2/3$. A running constant
$\alpha_V({\textbf q}^2)$ controls the behavior of standard
perturbation theory (SPTh) at low momentum scales. The formal
expression for $V_C(r)$ in position space can be written as
\begin{equation}
\label{eq:position space}
V_C(r)\,=\,-\,C_F\,\frac{\alpha_s(r)}{r}\,,
\end{equation}
where
\begin{equation}
\label{eq:alpha(r)}
\alpha_s(r)\,=\,\frac{2}{\pi}\,\int\limits_0^{\infty}dq\,\,\frac{\sin\,qr}{q}\,\,\alpha_V({\textbf
q}^2).
\end{equation}
Up to two loops
\begin{equation}
\label{eq:alpha_V}
 \alpha_V({\textbf
q}^2)\,=\,\frac{4\pi}{\beta_0\,t}\left(1\,-\,\frac{\beta_1}{\beta_0\,^2}\,\frac{\ln
t}{t}\right),
\end{equation}
where $\beta_i$ are the coefficients of the QCD $\beta$-function,
\begin{equation}
\beta_0\,=\,11\,-\frac{2}{3}\,n_f,\,\,\,\,\,\,\,\beta_1\,=\,102\,-\,\frac{38}{3}\,n_f,
\end{equation}
and
\begin{equation}
\label{eq:t} t\,=\,\ln\,\frac{{\textbf q}^2}{\Lambda_V^2}
\end{equation}
In what follows we use $n_f\,=\,3$.

 The conventional  coupling (\ref{eq:alpha_V})
is analytically singular at a scale ${\textbf
q}^2\,=\,\Lambda_{V}^2$, so the SPTh itself is not well defined in
the infrared domain where the coupling become large. This problem
can be traced back to the fact that the integral over the running
coupling which appears in Eq.(\ref{eq:alpha(r)}) is ill defined.
As a result, $\alpha_s(r)$ is known only in the perturbative
region, $r\,\lesssim\, 0.1$ fm. Estimates of the average
interquark distances in light baryons are in the vicinity of 0.7
fm which is certainly outside of the perturbative region.

There also exists the possibility of defining a running coupling
which stays finite in the infrared. One such example is the
``timelike'' effective coupling which is used in the dispersive
approach \cite{Ball:1995}. The idea is that such a coupling may
give an effective measure of interaction at low scale
\cite{Dokshitzer:1995}. An alternative procedure is to define the
fundamental coupling in QCD from a given physical observable
\cite{Brodsky:2003}. For our purposes, we find it convenient, as a
useful approximation, to define the strong coupling constant
$\alpha_B(r)$ in the BPTh \cite{Simonov1995}. In momentum space,
\be\label{eq:alpha_B}\alpha_B({\textbf q}^2)\,=\,\alpha_s({\textbf
q}^2\,+\,m_B^2),\ee where $m_B\,\sim\,1$ Gev is the appropriate
mass parameter~\footnote{~This parameter has the meaning of being
the lowest hybrid excitation, $m_B=M(Q{\bar Q}gg)-M(Q{\bar Q}g)$;
from comparison with the lattice static potential $m_B\,\sim\,1\,$
GeV.}. The logic behind Eq. (\ref{eq:alpha_B}) is that
the perturbative gluon propagator is modified strongly at $q\,\lesssim
m_B$ by the physics of large distances. This result can be
conventionally viewed as arising from the interaction of a gluon
with background vacuum fields.

 We define $\alpha_B(r)$, as well as the Coulomb--like potential in the
configuration  space via expressions similar to (\ref{eq:position
space}) and (\ref{eq:alpha(r)})
\begin{equation}
\label{eq:alpha_B(r)}
\alpha_B(r)\,=\,\frac{2}{\pi}\,\int\limits_0^{\infty}dq\,\,\frac{\sin\,qr}{q}\,\,\alpha_B({\textbf
q}^2),
\end{equation}
\begin{equation}
V_C(r)\,=\,-\,C_F\,\frac{\alpha_B(r)}{r}\,.
\end{equation}
For $\alpha_B({\textbf q}^2)$, we use the two--loop result
(\ref{eq:alpha_V}) with the substitution \begin{equation}
\label{eq:t} t\,\rightarrow\,t_B\,=\,\ln\,\frac{{\textbf
q}^2\,+\,m_B^2}{\Lambda_V^2}
\end{equation}
The resulting coupling $\alpha_B({\textbf q}^2)$ is finite in the
infrared, ${\textbf q}^2\to 0$. In the ultraviolet region,
${\textbf q}^2\gg m_B^2$ one recovers the SPTh result. In
configuration space the background coupling constant $\alpha_B(r)$
exists for all distances and saturates at some critical, or
freezing, value for $r\,\gg\, 1/m_B$. The specific choice of the
parameters in Eq. (\ref {eq:t}) will be discussed in Sec.
\ref{sect:results}.

\section{Hyperspherical formalism for three-quark systems.}
Following our previous analysis \cite{DNV} we use the
hyperspherical method to calculate the masses of the ground and
excited hyperon states. The idea of hyperspherical method is to generalize
the simplicity of the spherical harmonic expansion for the angular
functions of a two particle in three dimensional space to a system of N particles
by introducing in a 3N-3 dimensional space a global length $R$ called the hyperradius,
and a set of angles, $\Omega$. In this section, we briefly review the
hyperspherical formalism as applied to our specific problem.

Let us introduce in the system of three quarks with dynamical
masses $\mu_i$ and coordinates ${\bsl r}_i$ the three-body Jacobi
coordinates in the six-dimensional coordinate space as
\be\label{eq:Jacobi}
\bsl{\rho}_{ij}=\sqrt{\frac{\mu_{ij}}{\mu_0}}\,(\bsl{r}_i-\bsl{r}_j),\,\,\,\,\,\,
\bsl{\lambda}_{ij}=\sqrt{\frac{\mu_{ij,\,k}}{\mu_0}}
\left(\frac{\mu_i\bsl{r}_i+\mu_j\bsl{r}_j}{\mu_i+\mu_j}-\bsl{r}_k\right)\,,\ee
($i,j,k$ cyclic), where $\mu_{ij}$ and $\mu_{ij,k}$ are the
reduced dynamical quark masses:
\begin{equation}\mu_{ij}=\frac{\mu_i\mu_j}{\mu_i\,+\,\mu_j},~~~~
\mu_{ij,\,k}=\frac{(\mu_i\,+\,\mu_j)\mu_k}{\mu_i\,+\,\mu_j\,+\,\mu_k},\end{equation}
and $\mu_0$ is an arbitrary parameter with the dimension of mass,
which drops out in the final expressions. The baryon wave function
depends on the Jacobi coordinates (\ref{eq:Jacobi}). Now we switch
from Jacobi coordinates to hyperspherical coordinates in
coordinate space \bea
&&R^2\,=\,\bsl{\rho}^2+\bsl{\lambda}^2,\nonumber\\&&
\rho\,=\,R\,\sin\theta,\,\,\,\,\,
\lambda\,=\,R\,\cos\theta,\,\,\,\, 0 \le \theta \le \pi/2,\eea
where $R$ is the six-dimensional hyperradius that is invariant
under quark permutations. In what follows we omit the indices $i$
and $j$.

The explicit expression for the kinetic energy operator $H_0$ for three-quark systems in
the center--of--mass system is remarkably simple in hyperspherical
coordinates \be \label{H_0_jacobi} H_0\,
=\,-\,\,\frac{1}{2\mu_0}\left( \frac{\partial^2}{\partial
R^2}+\frac{5}{R}\frac{\partial}{\partial R}+
\frac{\bsl{L}^2(\Omega)}{R^2}\right). \ee  In Eq.
(\ref{H_0_jacobi}) $\Omega$ denotes the five angular coordinates
$\theta,\,{\bsl n_{\rho}},\,{\bsl n_{\lambda}}$, and
$\bsl{L}^2(\Omega)$ is a squared hyperangular momentum operator.
Its eigenfunctions (the hyperspherical harmonics) are defined by
\begin{equation}
\label{eq: eigenfunctions} {\bf L}^2(\Omega)\,Y_{[K]}(\theta,{\bf
n}_{\rho},{\bf n}_{\lambda})\,=\,-K(K+4)Y_{[K]}(\theta,{\bf
n}_{\rho},{\bf n}_{\lambda}),
\end{equation}
with $K$ being the grand orbital momentum.

The wave function $\psi(\bsl{\rho},\bsl{\lambda})$ is written in a
symbolic shorthand notation as
\be\psi(\bsl{\rho},\bsl{\lambda})=\sum\limits_{[K]}\psi_{[K]}(R)Y_{[K]}(\Omega),\label{eq:ss}\ee
where the set $[K]$ is defined by  the orbital momentum of the
state and the symmetry properties.

In what follows we use the ansatz $K\,=\,K_{\rm min}$, where
$K_{\rm min}\,=\,0$ for $L\,=\,0$ and  $K_{\rm min}\,=\,1$ for
$L\,=\,1$. The accuracy of this approximation has been discussed
in \cite{NSV}. Our task is then extremely simple in principle: we
have to choose a zero-order wave function corresponding to the
minimal $K$ for a given $L$. The corresponding hyperspherical
harmonics are \bea &&
Y_0\,=\,\sqrt{\frac{1}{\pi^3}}\,,\,\,\,\,K\,=\,0,\nonumber\\&&
\bsl{Y}_{\rho}\,=\,\sqrt{\frac{6}{\pi^3}}\,\frac{\bsl{\rho}}{R}\,,\,\,\,\,\,\,\,
\bsl{Y}_{\lambda}\,=\,\sqrt{\frac{6}{\pi^3}}\,\frac{\bsl{\lambda}}{R}\,,\,\,\,\,K\,=\,1.\eea

For $nns$ baryons we use the basis in which the strange quark is
singled out as quark $3$ but in which the nonstrange quarks are
still antisymmetrized. In the same way, for the $ssn$ baryon we
use the basis in which the nonstrange quark is singled out as
quark $3$.  The $nns$ basis states diagonalize the confinement
problem with eigenfunctions that correspond to separate
excitations of the nonstrange and strange quarks (${\rho}$\,- and
${\lambda}$\, excitations, respectively). In particular,
excitation of the $\bsl{\lambda}$ variable unlike excitation in
$\bsl{\rho}$ involves the excitation of the ``odd'' quark ($s$ for
$nns$ or $n$ for $ssn$). The nonsymmetrized $uds$ and $ssq$ bases
usually provide a much simplified picture of the states
\cite{CIK81}.

For the purpose of implementing our truncation scheme, we
introduce the reduced functions $u_{\nu}(R)$:
\be\label{eq:reduced}\Psi_{\nu}(R,\Omega)\,=\,\frac{u_{\nu}(R)}{R^{5/2}}\cdot{
Y}_{\nu}(\Omega), \ee where $\nu\,=\,0$~ for $L\,=\,0$,~and
$\nu\,=\,\rho,\,\lambda$ for $L\,=\,1$~\footnote{In what follows,
for ease of notation  we will drop the magnetic quantum numbers of
the vector spherical harmonics.}, as well as a new variable \be
x\,=\,\sqrt{\mu_0}\,R\,=\,
\left(\sum_i\,\frac{\mu_1\,\mu_2}{M}\,r_{12}^2\,+\,\frac{\mu_2\,\mu_3}{M}\,r_{23}^2\,+\,
\frac{\mu_3\,\mu_1}{M}\,r_{31}^2\right)^{1/2},\ee that does not
depend on $\mu_o$. Inserting Eq. (\ref{eq:reduced}) into the
Schr\"{o}dinger equation for $\Psi_{\nu}(R,\Omega)$ and averaging
the interaction $V=V_Y+ V_{C}$ over the six-dimensional sphere
$\Omega$ with the weight $|Y_{\nu}|^2$, one obtains the
one-dimensional Schr\"odinger equation for $u_{\nu}(x)$
 \be\label{eq:se}\frac{d^2
u_{\nu}(x)}{dx^2}\,+\,2\left(E_0\,-\,\frac{(K+\frac{3}{2})(K+\frac{5}{2})}{2\,x^2}\,-
\,V_{\rm Y}^{\nu}(x)\,-\,V_{\rm
C}^{\nu}(x)\right)u_{\nu}(x)\,=\,0,\ee where

\be\label{string} V_{\rm Y}^{\,\nu}(x)\,=\,\int
\,|Y_{\nu}\,(\theta,\chi)|^2\,V_{\rm Y}({\bf r}_1,\,{\bf
r}_2,\,{\bf r}_3)\,d\Omega\,=\,
\sigma\, b_{\nu}\,x,\ee and\be\label{Coulomb}V_{\rm
Coulomb}^{\,\nu}(x)\,=\,-\,\frac{2}{3}\,\int
\,|Y_{\nu}\,(\theta,\chi)|^2\,\sum_{i\,<\,j}\,\frac{\alpha_B(r_{ij})}{r_{ij}}\,\,\,d\Omega\,=\,
-\,\frac{2}{3}\,\frac{a_B^{\nu}(x)}{x}.\ee The constants $b_{\nu}$
in Eq. (\ref{string}) are defined by the two--dimensional
integrals in the plane $(\theta,\,
\cos\varphi\,=\,\bsl{\hat\rho}\,\bsl{\hat\lambda})$. Explicit
expressions for these integrals
 are written in the appendix of Ref. \cite{DNV}. The functions $a_B^{\nu}(x)$
 are

\begin{equation}
\label{a_B(x)}
a_B^{\nu}(x)\,=\,\sum_{i<j}\,\sqrt{\mu_{ij}}\,\int\, \,
\alpha_B\left(\frac{x\sin\theta}{\sqrt{\mu_{ij}}}\right)\frac{d\omega_{\nu}}{\sin\theta},
\end{equation}
where

\be d\omega_0\,=\,\frac{16}{\pi}\sin^2\theta\cos^2\theta\,
d\theta,\ee and \be d\omega_{\rho}\,=\,
\frac{32}{\pi}\,\sin^4\theta\cos^2\theta\,
d\theta,\,\,\,\,\,\,\,\,\,\,\,\, d\omega_{\lambda}\,=\,
\frac{32}{\pi}\,\sin^2\theta\cos^4\theta\,d\theta.\ee
\section{Results and discussion}
\label{sect:results} In the previous sections, we introduced the
Hamiltonian we use to obtain baryon spectra. This Hamiltonian
contains five parameters: the current quark masses $m_n$ and $m_s$,
the string tension $\sigma$, and two parameters $\Lambda_V$ and $m_B$
defining $\alpha_B({\bf q}^2)$ in Eq. (\ref{eq:t}). Let us
underline that they are not the fitting parameters. In our
calculations we use $\sigma\,=\,$0.15 GeV$^2$ found in the SU(3)
QCD lattice simulations \cite{Suganuma2003}. We employ the current
light quark masses ${m}_u\,=\,{ m}_d\,=\,7\,$ MeV and the bare
strange quark mass $m_s\,=\,$ 175 MeV found previously from the
fit to $D_s$ spectra \cite{KNS}. This value of the strange quark
mass is consistent with the QCD sum rules estimation
$m_s\,$(2~GeV)$\,=\,(125\pm40)$ MeV \cite{Leutwyler1996}. More
recent estimation yields
$m_s\,$(2~GeV)$\,=\,$$\,\,(90\pm10)\,$MeV~ \cite{Chetyrkin2006}.
Recall that in our Hamiltonian approach the current mass $m_s$
enters at a much lower scale. For the remaining two parameters
$\Lambda_V$ and $m_B$ in Eq. (\ref{eq:t}) we employ the values
\be\label{eq:parameters} \Lambda_V\,=\,(0.36\pm 0.02)\,{\rm
GeV},\,\,\,\,\, m_B\,=\,(1\pm0.05)\,{\rm GeV},\ee determined
previously in  Ref. \cite{BK2001}. The result is consistent with
the freezing of $\alpha_B(r)$
with a magnitude $\sim\,0.5-0.6$, see Table \ref{tab:saturation}.
The behavior of $\alpha_B(r)$ for $\Lambda_V\,=\,0.36$ GeV is
shown in Fig. 1 for three different values of $m_B$.  We use the
error bars in Eq. (\ref{eq:parameters}) to illustrate the
sensitivity of the baryon masses to the chosen input. Note that
$\alpha_B(r)$ increases with $\Lambda_V$ and, for fixed
$\Lambda_V$, decreases with $m_B$.

\begin{figure}
\begin{center}
\epsfxsize=7.5cm \epsfysize=7.5cm \epsfbox{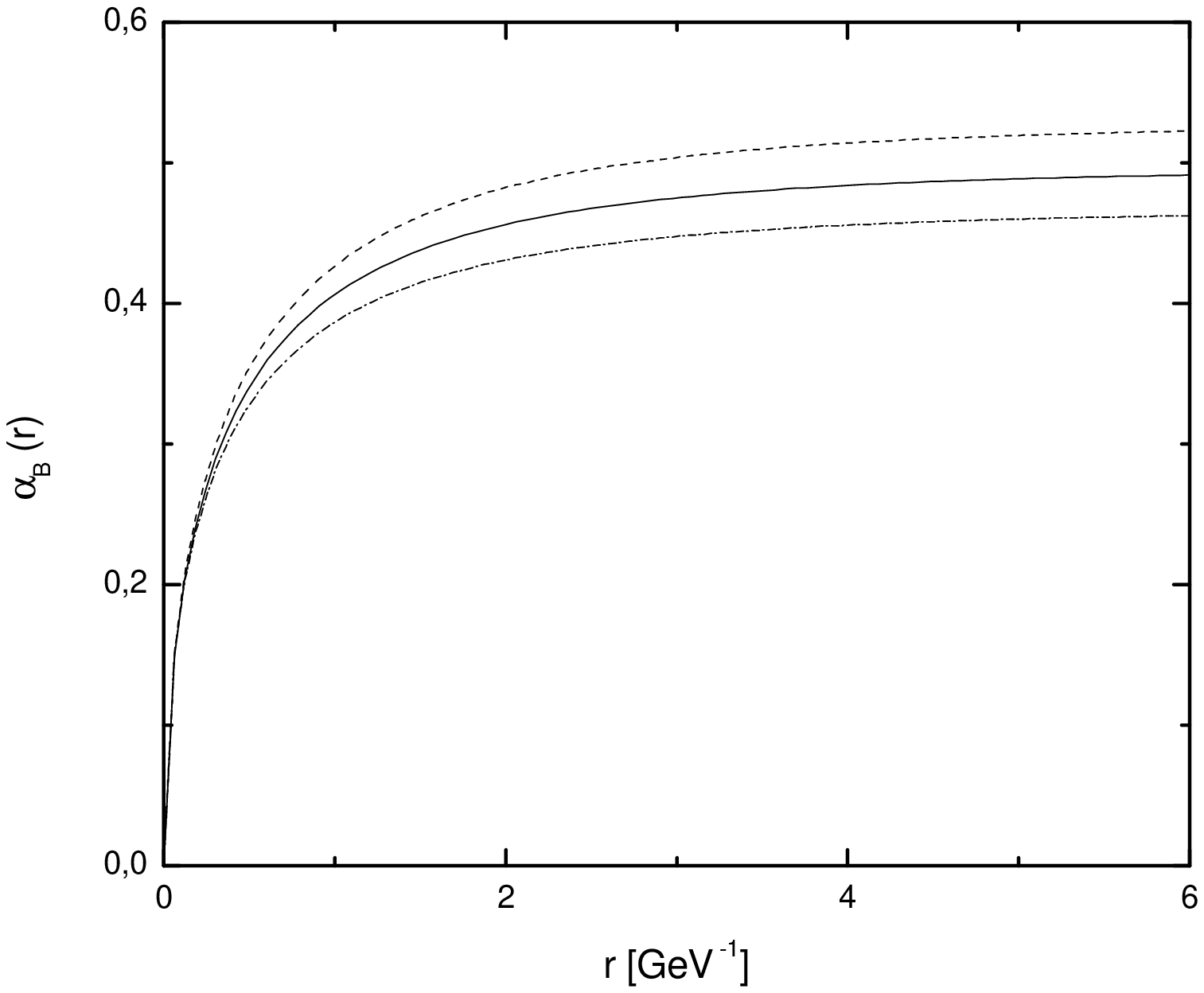}
\end{center}
\caption{The dependence of the running coupling constant $\alpha_B(r)$ on the distance for $m_B\,=\,$1 GeV and $\Lambda_V\,=\,$
0.36 (solid line), 0.38 (dashed line) and 0.34 GeV
(dotted--dashed line)}
\end{figure}

\begin{table}[t]
 \caption{The freezing values of $\alpha_B(\infty)$ for the different choices of $\Lambda_V$ and $m_B$.
 The values of $\Lambda_V$ and $m_B$ are given in GeV.} \vspace{5mm}
\centering
\begin{tabular}{c|ccccccccc}\hline\hline\\
$\Lambda_V$&&0.34&&&0.36&&&0.38\\\\ $m_B$&~~~~~~0.95&1.00&1.05&~~~~~~0.95&1.00&1.05&~~~~~~0.95&1.00&1.05\\
\\$\alpha_B({\infty})$&~~~~~~0.492&0.467&0.445&~~~~~~0.526&0.496&0.471&
~~~~~~0.563&0.528&0.500\\ \\
\hline\hline
\end{tabular}
\label{tab:saturation}
\end{table}

We begin the discussion of our results by examining the
predictions for the ground states of  the $nnn$, $nns$ and $snn$
baryon with $L\,=\,0$. In Table \ref{tab:L=0} are displayed the
$nnn$, $nns$ and $snn$ masses for the three choices of
$\Lambda_V$: 0.34, 0.36 and 0.38 GeV, and three choices of $m_B$:
0.95, 1.00 and 1.05 GeV. In this table we also show the dynamical
masses $\mu_i$  found from the minimum condition (\ref{eq:mc}).

The
results in Table \ref{tab:L=0} show that the baryon masses decrease when $\Lambda_V$ increases and, for
fixed $\Lambda_V$, the baryon masses decrease when $m_B$
increases. This comes as no surprise: the effect can be easily
read off from the results of Table \ref{tab:saturation} and Fig.
1. Increasing $\Lambda_V$ for fixed $m_B$ and decreasing $m_B$ for fixed $\Lambda_V$
leads to increased running constant $\alpha_B(r)$ and
respectively, to decreased baryon masses.
Upon varying the parameters $\Lambda_V$ and $m_B$, we
obtain the baryon masses in the interval 1161\,--\,1209 MeV
($nnn$), 1246\,--\,1297 MeV ($nns$), and 1330\,--\,1383 MeV
($ssn$). The difference in the mass values is mostly due to the
difference in the running of the strong coupling in the
midmomentum regime.
\begin{table}[t]
 \caption{The masses of the $nnn,~nns~and~ssn$ baryons with $L\,=\,0$.
 All quantities are given in MeV.}\vspace{5mm}

\label{tab:baryon_masses} \centering
\begin{tabular}{ccccccccccc} \hline\hline\\
&&&~~$nnn$&&& ~~$nns$&&&~~$ssn$&
\\  \\\hline\hline\\
$\Lambda_V$~~~&~~$m_B$~~&$\mu_1$&$\mu_3$~&$M_B$~~&$\mu_1$&$\mu_3$~&$M_B$&~~$\mu_1$&$\mu_3$~&$M_B$\\
\\ \hline\\
340&~~1050~~&~~407~~&~~407~~&1209&412&452&1297&457&417&1383\\
&~~1000~~&~~409~~&~~409~~&1201&414&453 &1288&458&419&1373\\
&~~950~~&~~410~~&~~410~~&1190&415&455&1277&460&$421$&1363\\ \\\hline\\
360&~~1050~~&~~409~~&~~409~~&1197&415&454&1286&459&420&1371\\
&~~1000~~&~~411~~&~~411~~&1187&417&456 &1276&461&422&1360\\
&~~950~~&~~414~~&~~414~~&1177&419&458&1263&463&$424$&1347\\ \\\hline\\
380&~~1050~~&~~412~~&~~412~~&1186&418&457&1274&462&423&1358\\
&~~1000~~&~~414~~&~~414~~&1174&420&459 &1252&464&425&1345\\
&~~950~~&~~417~~&~~417~~&1161&422&461&1246&466&$428$&1330\\ \\\hline\\
\end{tabular}
\label{tab:L=0}
 \vspace{1mm}

\end{table}
\begin{table}[t]
 \caption{Comparison of the baryon masses $M_0\,+\,C$ in
Eq. (\ref{M_B}) calculated
 using the RCC $\alpha_B(r)$ without taking into account the string correction with those obtained for
the FCC $\alpha_s\,=\,0.39$ from Ref.
\cite{NSV}. The quantities shown in the rows labeled by RCC
 have been calculated for $\Lambda_V\,=\,360$ MeV, $m_B\,=\,1000$
 MeV, the upper error bars correspond to $\Lambda_V\,=\,340$ MeV,
 and the lower error bars correspond to $\Lambda_V\,=\,380$ MeV. The masses of the
 baryons and dynamical masses $\mu_i$ are given in MeV.}
 \vspace{5mm}

\centering\label{tab:baryon_masses}

\begin{tabular}{ccccccccccc} \hline\hline\\
&&&$nnn$&&& $nns$&&&$ssn$
\\  \\\hline\hline\\
&${\bf L}$&$\mu_1$&$\mu_3$~&$M_0\,+\,C$~~&$\mu_1$&$\mu_3$~&$M_0\,+\,C$&~~$\mu_1$&$\mu_3$~&$M_0\,+\,C$\\
\\ \hline\\
RCC&${\bf
0}$&411$^{\,+6}_{\,-4}$&411$^{\,+6}_{\,-4}$&1187$^{\,+22}_{\,-26}$&417$^{+5}_{-5}$&456$^{+5}_{-4}$&1276$^{+22}_{-30}$
&461$^{+5}_{-5}$&422$^{+6}_{-5}$&1360$^{+23}_{-30}$\\ \\
FCC&&408&408&1209&414&453&1298&458&419&1384\\ \\\hline\\ RCC&${\bf
1}_{\rho}$&454$^{+2}_{-2}$&454$^{+2}_{-2}$&1695$^{+9}_{-10}$&477$^{+2}_{-2}$
&460$^{+2}_{-2}$&1774$^{+9}_{-10}$&516$^{+4}_{-2}$&424$^{+3}_{-2}$&1832$^{+9}_{-10}$\\
\\
FCC&&457&457&1674&482&459&1751&520&424&1810\\ \\ \hline\\RCC&${\bf
1}_{\lambda}$&469$^{+4}_{-3}$&469$^{+4}_{-3}$&1625$^{+13}_{-15}$&456$^{+2}_{-4}$&540$^{+2}_{-3}$&
1687$^{+13}_{-15}$&496$^{+3}_{-3}$&513$^{+3}_{-3}$&1771$^{+14}_{-15}$
\\ \\
FCC&&457&457&1674&441&534&1738&483&506&1827\\ \\ \hline\hline
\end{tabular}
\label{tab:comparison}
\end{table}

\begin{table}[t]

\caption{Masses of the $\rho$ and $\lambda$ baryon excitations for
$m_B\,=\,$1000 MeV taking into account the string
corrections, Eqs. (\ref{eq:rho}), (\ref{eq:lambda}). The masses of
the baryons, string corrections and dynamical masses $\mu_i$ are
given in MeV.}
 \vspace{5mm}

\centering
\begin{tabular}{cccccccccc} \hline\hline\\
Baryon&Excitation&$\Lambda_V$& $\mu_1\,=\,\mu_2$ & $\mu_3$& $M_0\,+\,C$&$\Delta M_{\rm string}$&$M_B$\\
\\ \hline\\
$nnn$&${\bsl 1}_{\rho}$&340&452&452&1704&-60&1644\\
 &&360&454
&454&1695&-59&1636 \\
&&380&456&456&1685&-59&1626\\ \\ \hline\\
&${\bsl 1}_{\lambda}$&340&466&466&1638&-58&1580\\
&&360&469&469&1625&-58&1567 \\
&&380&471&471&1610&-57&1553\\ \\
 \hline\hline\\
 $nns$&${\bsl 1}_{\rho}$&340&475&458&1783&-55&1728\\
 &&360&477&460&1774&-55&1719\\
 &&380&479&462&1764&-55&1709\\ \\
 \hline\\
  $nns$&${\bsl 1}_{\lambda}$&340&452&537&1700&-51&1649\\
  &&360&455&540&1687&-51&1636\\
 &&380&458&542&1672&-51&1621\\ \\
 \hline\hline\\

$ssq$&${\bsl 1}_{\rho}$&0.34&   514 & 422  &
1841&-48&1793\\
&& 0.36&516&424&1832&-48 &1784\\
&&0.38&518&427&1822&-48&1774\\
\\ \hline \\
&${\bsl 1}_{\lambda}$&0.34&493&510&1785&-51&1734\\
&&0.36&496&513&1771&-51&1720\\  &&0.38&499&516&1756&-51&1705\\
\\\hline\hline

\end{tabular}
\label{tab:P--wave}
\end{table}
\begin{table*}[t]
\caption{Low-lying $\Xi$  spectrum of spin $L\,=\,1$ predicted by
the nonrelativistic quark models of Chao, Isgur and Karl
\cite{CIK81} and of Pervin and Roberts \cite{PR}, the relativized
quark model of Capstick and Isgur \cite{CI86}, the Glozman-Riska model
\cite{GR96}, the large $N_c$ analysis \cite{CC00}, the algebraic model
\cite{BIL00}, QCD sum rules \cite{LL02}, and the Skyrme model
\cite{Oh07}. The question mark in the last column means that the
$J^P$ quantum numbers are not identified by PDG. The masses are
given in MeV.} \vspace{2mm}

\centering
\begin{tabular}{c|cccccccccc} \hline\hline\\
State & \cite{CIK81} & \cite{PR}&\cite{CI86} & \cite{GR96} &
\cite{CC00} & \cite{BIL00} &\cite{LL02}&\cite{Oh07}&This work&PDG

\\ \\ \hline \\ \\
$\Xi(\frac12^-)$ & $1785$ & 1725&$1755$ & $1758$ & $1780$ & $1869$
&
$1550$&1660&$1720^{+14}_{-15}$& $\Xi(1690)\,?$\\
$\Xi(\frac32^-)$ & $1800$ & $1759$&$1785$ & $1758$ & $1815$ &
$1828$ & $1840$ &1820&$1720^{+14}_{-15}$&$\Xi(1820)\,$\\ \\
\hline\hline
\end{tabular}

\label{tab:xi theory}
\end{table*}
It is  instructive to compare the results obtained for the running
coupling constant (RCC) with those obtained for the freezing
coupling constant (FCC) $\alpha_s^{(0)}\,=\,0.39$ \cite{DNV}. In
Table \ref{tab:comparison} such comparison is presented for
the $S$-- and $P$--baryon states for the particular set of the
parameters $\Lambda_V\,=\,(0.36\,\pm\,0.02)$ GeV and $m_B\,=\,1$
GeV. In this table we compare the dynamical masses $\mu_i$ and the
baryon masses $M_0\,+\,C$ in Eq. (\ref{M_B}).
The quantities shown in the rows labeled by RCC
 have been calculated for  $m_B\,=\,1.00$
 GeV, the central results correspond to $\Lambda_V\,=\,0.36$ GeV,
 the upper error bars correspond to $\Lambda_V\,=\,0.34$ GeV, and
 the lower error bars correspond to $\Lambda_V\,=\,0.38$ GeV.

 From Table \ref{tab:comparison}, it can be seen that
for the $S$--states the baryon masses calculated for the RCC and
FCC are all in agreement with each other
within the error bars, although the central
values for the $nnn$, $nns$ and $ssn$ states are typically lower
by about 20 MeV than those for the FCC. For the $P$--waves, the
central values of the masses of $\rho$ excitations are typically
20 MeV heavier than those for the FCC while the central mass
values of the $\lambda$ excitations are smaller by about 50 MeV
than those values calculated for $\alpha_s^{(0)}\,=\,0.39$. Note
that introducing the RCC
eliminates the
degeneracy of the $\rho$ and $\lambda$--excitations for the $nnn$
states found previously for the FCC \cite{NSV}.

Now let us consider the string
 correction for the confinement potential for the orbital excitations. Explicit
 calculation \cite{DN}
 of the matrix elements in Eq. (\ref{eq:Delta_M}) yields
\be\label{eq:rho}\Delta M^{\rho}_{\rm
string}\,=\,-\,\frac{64\,\sigma}{45\pi}\,\cdot\,\frac{1}{\mu^{3/2}}\,\sqrt{\frac{1\,+\,\kappa}{2\,+\,\kappa}}\,
\,\gamma_{\rho},\ee
 \be\label{eq:lambda}\Delta
 M^{\lambda}_{\rm string}=-\frac{64\sigma}{45\pi}\,\cdot\,\frac{\kappa}{\mu^{3/2}(2\,+\kappa)^{3/2}}
 \left(
\frac{1}{\kappa^{5/2}}\,+\,2\,\sqrt{1\,+\,\kappa}
\right)\,\gamma_{\lambda}, \ee
 where $
\mu_1\,=\,\mu_2\,=\mu$, $\kappa\,=\,\mu_3/\mu$,
 \be\label{eq:gamma_nu}
\gamma_{\nu}\,=\,\int\limits_0^{\infty}
 \frac{{
 u}_{\nu}^2(x)}{x}dx,\,\,\,\,\,\,\,\nu=\,\rho,\,\lambda,
 \ee
and the wave functions $u_{\nu}(x)$ are normalized to unity.

The wave function corrections $\gamma_{\nu}$ in
(\ref{eq:gamma_nu}) which influence the string correction does not
essentially depend  on the baryon flavor nor on the type of
excitation. {\it E.g.} for $\Lambda_V\,=\,$ 0.36 GeV and
$m_B\,=\,1$ GeV we get (in units of GeV$^{1/2}$)
$$\gamma_{\rho}\,=\,0.328\,(nnn),\,\,\,\,\,\,\gamma_{\rho}\,=\,0.326\,(nns),
\,\,\,\,\,\,\gamma_{\rho}\,=\,0.325\,(ssn),$$ and
$$\gamma_{\lambda}\,=\,0.333\,(nnn),\,\,\,\,\,\,\gamma_{\lambda}\,=\,0.332\,(nns),
\,\,\,\,\,\,\gamma_{\lambda}\,=\,0.331\,(ssn).$$

The results for the $P$--wave excitations of the $nnn$, $nns$ and
$ssn$ baryons when are taking into account the string correction
along with the values for the string correction itself are shown
in Table \ref{tab:P--wave}. Analysis of the data shows that string
correction does not essentially depend  on the baryon flavor nor
on the type of excitation. As a result of the weak dependence of
the string correction on a baryon flavor and the type of
excitation the masses for all considered baryons become smaller by
about the same value $\sim\,50-60$ MeV (see the 6$^{\rm th}$
column of Table \ref{tab:P--wave}).

The baryon energies agree reasonably well with the Particle Data Group (PDG) listing
particularly if we take into consideration that spin interactions
are neglected. For instance, for $L\,=\,0$ we get
$\frac{1}{4}(\Lambda\,+\,\Sigma\,+\,2\,\Sigma^*)_{\rm
theory}\,=\,$ 1276$^{+23}_{-30}$ MeV, where  the error bars
correspond to the variation of the hyperon masses within the
chosen range of $\Lambda_V$ and $m_B$ versus
$\frac{1}{4}(\Lambda\,+\,\Sigma\,+\,2\,\Sigma^*)_{\rm exp}\,=\,$
1267 MeV. For the $\Xi$, we have $\Xi_{\rm
theory}\,=\,$1360$^{+23}_{-30}$ MeV, whereas $\Xi_{\rm
exp}\,=\,$1315 MeV. However, for the nucleon we get
$\frac{1}{2}\,(N\,+\,\Delta)_{\rm theory}\,=\,$ 1187$^{+22}_{-21}$
MeV, which is
about 100 MeV heavier than $\frac{1}{2}\,(N\,+\,\Delta)_{\rm
exp}\,=\,$ 1085 MeV. The difference can be ascribed to the effects
of spin--dependent quark--quark interactions modeled after the
effect of gluon exchange in QCD \cite{DeRujula}  or arising from
one--boson exchange \cite{GR96} that are completely omitted in the
present approach~\footnote{Note that the one--boson exchange
effects are important for baryons containing the scalar $nn$
diquarks ($N$ and $\Lambda$) and much less important for the
baryons containing the axial diquarks ($\Delta$ and $\Sigma$), see
{\it e.g.} Ref. \cite{VGV}.}. Another source of the discrepancy is
the systematic error associated with the use of the AF formalism,
which is maximal for the $S$--wave $nnn$ states \cite{NSV}.

The physical $P$-wave states are not pure $\rho$ or $\lambda$
excitations but  linear combinations of all states with a given
total momentum $J$. Most physical states are, however, close to
pure $\rho$ or $\lambda$ states
\cite{CIK81}. For example, the masses of N(1535) and N(1520)
resonances with $J^P\,=\,\frac{1}{2}^-$ and
$J^P\,=\,\frac{3}{2}^-$, respectively, both match with the mass of
the $\lambda$--excitation for the $nnn$ baryon from Table
\ref{tab:P--wave}: $M_{\lambda}(nnn)\,=\,1567^{+13}_{-14}$ MeV.
The masses of $\Sigma(1620)$ and $\Sigma(1670)$ states with
$J^P\,=\,\frac{1}{2}^-$ and $J^P\,=\,\frac{3}{2}^-$, respectively,
match very closely with the mass of the $\lambda$--excitation for
the $nns$ baryon: $M_{\lambda}(nns)\,=\,1636^{+13}_{-15}$ MeV. The
masses of the $\rho$--excitations which correspond to $P$--states
of the light diquarks are typically 60 - 80 MeV higher.

The comparison of  our result for the negative parity ground state
in the $\Xi$ channel, $M_{\lambda}(ssn)\,=\,1720^{+14}_{-15}$ MeV
with the other theoretical predictions for this state and the
result from the PDG listing is presented in Table \ref{tab:xi
theory}.
\section{Conclusions}
In this paper the study of the ground and excited states of the
$qqq$, $qqs$ and $ssq$ baryons by including into the EH derived
using the FCM the effects of the running strong coupling constant
$\alpha_B(r_{ij})$ in the perturbative Coulomb--like part of the
three--quark potential is presented. The results have refined our
previous studies
obtained for the freezing coupling constant.
The three-quark problem has been solved using the hyperspherical
approach. The string correction for the confinement potential of
the orbitally excited baryons is estimated. For each baryon, we
have calculated the dynamical quark masses $\mu_i$ from Eq.
(\ref{eq:mc}) and the baryon masses (\ref{M_B}) with the
self-energy corrections (\ref{self_energy}) and the string
corrections (\ref{eq:rho}), (\ref{eq:lambda}). The main results
are given in Tables \ref{tab:L=0}~-~\ref{tab:P--wave}. Our study shows
that a fairly good description of the $S$ and $P$--wave baryons
can be obtained with spin independent energy eigenvalues
corresponding to the confining along with Coulomb potentials. We
emphasize that no fitting parameter were used in our calculations.
This comparative study provides deeper insight into the quark
model results for which the constituent masses encode the QCD
dynamics. \vspace{3mm}

This work was supported in part by RFBR Grants No. 06-02-17120,
 No.
08-02-00657, No. 08-02-00677, and by the grant for scientific schools
NSh.4961.2008.2 .

\end{document}